\documentclass[useAMS,usenatbib]{lib/mnras}

\usepackage{natbib}
\usepackage{amssymb}
\usepackage{epsfig} 
\usepackage{scrtime}

\newcommand{\hb}{\mathrm{H}\beta}

\newcommand{\hii}{H\,\textsc{ii}}

\newcommand{\lsig}{$L(\mathrm{H}x) - \sigma$}
\newcommand{\lsigG}{$L - \sigma$}
\newcommand{\lsigb}{$L(\mathrm{H}\beta) - \sigma$}

\newcommand{\mincir}{\raise-3.truept\hbox{\rlap{\hbox{$\sim$}}\raise4.truept\hbox{$<$}\ }}

\title[Constraining the Dark Energy EoS with HII Galaxies]{Constraining the Dark Energy Equation of State with HII Galaxies}
\author[R.~Ch{\'a}vez et al.]
{ R.~Ch{\'a}vez,$^{1, 2, 3}$\thanks{E-mail:ricardo.chavez@mrao.cam.ac.uk} 
  M.~Plionis,$^{4, 5}$ S.~Basilakos,$^{6}$ R.~Terlevich,$^{1,7}$ E.~Terlevich,$^{1}$  
  \newauthor\  J.~Melnick,$^{8}$ F.~Bresolin,$^{9}$ and A.L. Gonz\'alez-Mor\'an$^{1}$ \\
$^{1}$Instituto Nacional de Astrof{\'i}sica {\'O}ptica y
Electr{\'o}nica, AP 51 y 216, 72000, Puebla, M{\'e}xico\\
$^{2}$Cavendish Laboratory, University of Cambridge, 19 J. J. Thomson Ave, Cambridge CB3 0HE, UK\\
$^{3}$Kavli Institute for Cosmology, University of Cambridge, Madingley Road, Cambridge CB3 0HA, UK\\
$^{4}$Physics Dept., Aristotle Univ. of Thessaloniki, Thessaloniki 54124, Greece \\
$^{5}$National Observatory of Athens, P.Pendeli, Athens, Greece \\
$^{6}$Academy of Athens, Research center for Astronomy and Applied
Mathematics, Soranou Efesiou 4, 11527, Athens, Greece \\
$^{7}$Institute of Astronomy, University of Cambridge, Madingley Road,
Cambridge CB3 0HA, UK\\
$^{8}$European Southern Observatory, Alonso de Cordova 3107, Santiago, Chile\\
$^{9}$Institute for Astronomy of the University of Hawaii, 2680
Woodlawn Drive, 96822 Honolulu,HI USA\\
}

\begin{document}

\date{v18 --- Compiled at \thistime\ hrs  on \today\  }

\pagerange{\pageref{firstpage}--\pageref{lastpage}} \pubyear{2016}

\maketitle

\label{firstpage}

\begin{abstract}
We use the HII galaxies \lsigG\ relation 
and the resulting Hubble expansion cosmological probe of a
sample of just 25 high-$z$ (up to $z \sim 2.33$) \hii\ galaxies, in
a joint likelihood analysis with other well tested cosmological probes
(CMB, BAOs) in an attempt to constrain the dark energy equation of
state (EoS). The constraints, although still weak, are in excellent
agreement with those of
a similar joint analysis using the well established SNIa Hubble
expansion probe. Interestingly, even with the current small
  number of available high redshift HII galaxies, the HII/BAO/CMB joint analysis gives a
13\% improvement of the quintessence dark energy 
cosmological constraints compared to the BAO/CMB joint analysis.

We have further performed extensive Monte Carlo simulations, with a
realistic redshift sampling, to explore the extent to which 
the use of the \lsigG\ relation, observed in \hii\ galaxies,
can constrain effectively the parameter space of the dark energy EoS. 
The simulations  predict substantial improvement in the constraints when
increasing the sample of high-$z$ \hii\ galaxies to 500, a goal that
can be achieved in reasonable observing times with existing large
telescopes and state-of-the-art instrumentation.

\end{abstract}


\section{Introduction}
The observational evidence for an accelerated cosmic expansion was first given by 
Type Ia Supernovae (SNIa) \citep{Riess1998, Perlmutter1999}. Since then, measurements of the
cosmic microwave background (CMB) anisotropies
\citep[e.g.][]{Jaffe2001, Pryke2002, Spergel2007, Planck2015} and of
Baryon Acoustic Oscillations (BAOs) \citep[e.g.][]{Eisenstein2005}, in
combination with independent Hubble parameter measurements
\citep[e.g.][]{Freedman2012}, have provided ample evidence of the
presence of a dark energy (DE) component in the Universe.

To the present day, the main geometrical tracer of the cosmic
acceleration has been SNIa  at redshifts $z \lesssim 1.5$
\citep[e.g.][]{Suzuki2012, Betoule2014}. It is of great
importance to use alternative geometrical probes at higher redshifts 
in order to verify the SNIa results and to obtain
more stringent constrains in the cosmological parameters solution
space \citep{Plionis2011}, with the final aim of discriminating among the various theoretical
alternatives that attempt to explain the accelerated expansion of the
Universe \citep[cf.][]{Suyu2012}. 

The \lsig\ relation between the velocity dispersion ($\sigma$) and
Balmer-line luminosity ($L[\mathrm{H}x]$, usually H$\beta$) of \hii\ galaxies has
already proven its potential as a cosmological tracer
\citep[e.g.][and references therein]{Melnick2000, Siegel2005, Plionis2011,
Chavez2012, Chavez2014, Terlevich2015}. It has been shown that the
\lsigb\ relation can be used in the local Universe to constrain the
value of $H_0$ \citep{Chavez2012}. At high-$z$ it can set constraints
on the parameters of the DE Equation of State (EoS) \citep{Terlevich2015}. 

\hii\ Galaxies are a  promising  tracer for the parameters of
the DE EoS precisely because they can be observed, using the current available
infrared instrumentation, up to $z \sim 3.5$ \citep[cf.][]{Terlevich2015}. Even when their scatter 
on the Hubble diagram is about a factor of two
larger than in the case of SNIa, this disadvantage is
compensated by the fact that  \hii\ galaxies  are observed to much larger 
redshifts than SNIa where  the degeneracies for different
DE models are substantially reduced \citep[cf.][]{Plionis2011}.

In addition, because the \lsigb\ relation systematic uncertainty sources\citep{Chavez2012, Chavez2014}  
are not the same as those of SNIa,  \hii\ galaxies constitute
an important complement to SNIa in the local
Universe, contributing to a better understanding of the systematic
errors of both empirical methods. 

In this paper we perform an HII/BAO/CMB joint likelihood analysis and
  compare the resulting cosmological constraints with those of
  a BAO/CMB and a SNIa/BAO/CMB joint likelihood analysis (for the
  latter we use the \emph{Union 2.1} SNIa compilation \citep{Suzuki2012}).

Furthermore, we present extensive Monte-Carlo simulations,
tailored to the specific uncertainties of the \hii\ galaxies
  \lsigb\ relation and currently available instrumentation, to
demonstrate its potential possibilities as a cosmological tracer 
to $z\lesssim 3.5$, to
probe a region where the Hubble function is very sensitive to the
variations of  cosmological parameters \citep{Melnick2000,
  Plionis2011}.

The paper is organised as follows: in section 2 we succinctly describe the data used 
and  associated systematic uncertainties; cosmological constraints
that can be obtained from the data are explored in section 3;  in
section 4 we discuss the Monte-Carlo simulations, in section
5 we discuss the  planned data acquisition in order  to
obtain better constraints on the cosmological parameters.
 Finally in section 6 we present our 
conclusions.

\section{H II galaxies data}
Our current sample  consists of a  low-$z$ subsample
 of 107 \hii\ galaxies ($0.01 \leq z \leq 0.16$)  extensively analysed in 
\citet{Chavez2014} and  24 Giant Extragalactic \hii\ Regions (GEHR) at
$ z \leq 0.01$  described in   \citet{Chavez2012}.
The sample also includes a high-$z$ subsample  composed by 6 star-forming
galaxies, selected from \citet{Hoyos2005, Erb2006b, Erb2006} and
\citet{Matsuda2011}, that we observed \citep{Terlevich2015} using
X-SHOOTER \citep{Vernet2011} at the Very Large Telescope in Paranal. The data of 
19 objects  taken from
\citet{Erb2006, Maseda2014} and \citet{Masters2014} complete the
sample. Altogether, the redshift
range covered by the high-$z$ subsample is 
$0.64 \leq z \leq 2.33$. 

It has been demonstrated  \citep[cf.][]{Terlevich1981, Melnick1988,
  Terlevich2003, Plionis2011, Chavez2012, Chavez2014} that the
\lsigb\ relation for \hii\ galaxies and GEHR can be used to measure
distances via the determination of their Balmer emission line
luminosity, 
$L(\mathrm{H}\beta)$,  and the velocity dispersion ($\sigma$) of the young starforming 
cluster from
measurements of the line width. The relevant relation can be expressed as:
\begin{equation}
	\log L(\mathrm{H}\beta) = (5.05 \pm 0.097) \log \sigma (\mathrm{H}\beta) + (33.11 \pm 0.145) .
\label{eq:LS}	
\end{equation}
Distance moduli are then obtained from:
\begin{equation}
\mu^{o} = 2.5 \log  L(\mathrm{H}\beta)_{\sigma} - 2.5 \log f(\mathrm{H}\beta) - 100.195
\label{eq:mu1}	
\end{equation}
where   $L(\mathrm{H}\beta)_{\sigma}$ is the luminosity estimated
from the \lsigb\ relation as in eq. (\ref{eq:LS}) and $f(\mathrm{H}\beta)$  is the measured flux in the H$\beta$ line. The uncertainty  on
the distance moduli, $\sigma_{\mu^{o}}$, is propagated from the
uncertainties in $\sigma_i$ and $f_i$ and the slope and intercept of
the distance estimator in eq. (\ref{eq:LS}).

\subsection{Systematic Errors}
\begin{table}
\center{
  \caption {Systematic error budget on the distance moduli, $\mu$. The typical uncertainty
    contribution of each source of systematic error is given. }
\tabcolsep 15pt
\begin{tabular}[h] { l l l }
\hline
\hline
Source & Error  \\
\hline
 Size of the Burst & 0.175 \\
 Age of the burst & 0.05 \\ 
 Abundances & 0.05 \\
 Extinction &  0.175 \\
\hline
 Total & 0.257 \\
\hline

\label{tab:SEB}
\end{tabular}
}

\end{table}
\subsubsection{Size of the burst}

The scatter found in the \lsigb\ relation for
\hii\ galaxies suggests a dependence on a second parameter  
\citep[cf.][]{Terlevich1981, Melnick1987}. Indeed
\citet{Chavez2014}, using SDSS DR7 effective Petrosian radii,
corrected for seeing, for a sample of local \hii\ galaxies, found 
 the size of the starforming region to be this second parameter.

For the high-$z$ samples, unfortunately, we do not have any
size measurements, so using it as a second parameter in the
correlation is impossible. The error induced by not using the size of
the burst as a second parameter appears in the uncertainties in the
slope and zero point of the \lsigb\ relation, i.e.\ our 
uncertainty values already incorporate this effect. In Table
\ref{tab:SEB} we show the typical contribution of the size of the
burst to the uncertainty on the distance moduli. 

\subsubsection{Age of the burst} 
\citet{Melnick2000} have demonstrated that \hii\ galaxies with equivalent width of H$\beta$, $W(\hb)
< 25\ \mathrm{\AA}$, do follow an \lsigb\ relation with a similar slope
but  different intercept than those with larger $W(\hb)$, 
i.e.\ older starbursts follow a parallel less luminous
\lsigb\ relation. For the study presented here, the starburst age is a
controlled parameter in the sense that we have selected our sample to
be composed of very young objects \citep[$\lesssim 5$ Myr, for 
instantaneous burst models cf.][]{Leitherer1999} by putting a high
lower limit to the value of $W(\mathrm{H}\beta) >
50\ \mathrm{\AA}$. Therefore only the youngest bursts were considered
and  in this way the effects of the age of the burst as a systematic
error on the \lsigb\ relation has been minimised; this selection also minimises
the contamination by an older underlying stellar component.

We have demonstrated \citep{Chavez2014}  that using the $W(\hb)$ as a
second parameter in the \lsigb\ correlation reduces only slightly the 
scatter because of the small dynamic  range of the age of our sample
objects. We chose  \citep{Terlevich2015}
 not to use the $W(\hb)$ as a parameter to
`correct' the \lsigb\ relation. Therefore, the small effect of the age
of the burst on the correlation manifests itself in the uncertainties of
the slope and zero point of the \lsigb\ relation that we are
adopting. In Table \ref{tab:SEB} we show the typical contribution of
the age of the burst to the uncertainty on the distance moduli.

\subsubsection{Abundances}
The oxygen abundance of \hii\ galaxies was considered in the past
\citep[eg.][]{Melnick1987, Melnick2000, Siegel2005} as a second
parameter for the \lsigb\ relation.
We have explored again this issue in 
\citet{Chavez2014}  for our local sample and
concluded that the effect albeit present is very small. 

We chose \citep{Terlevich2015} not to use the oxygen
abundance as a parameter to `correct' the \lsigb\ relation, and thus
the small effect of the metallicity of the burst on the correlation is
already part of the uncertainties of the slope and zero point of the
\lsigb\ relation that we are adopting. The typical contribution of the abundances to the uncertainty on the
distance moduli is shown in  Table \ref{tab:SEB}.
 
\subsubsection{Extinction}
The internal extinction correction was performed on the low-$z$
subsample following the procedure  described in \citet{Chavez2014} and
using the extinction coefficients derived from SDSS DR7 spectra. For
the high-$z$ subsample 
we  used the extinction coefficients given in the literature
\citep{Erb2006b, Erb2006, Matsuda2011, Maseda2014,
  Masters2014}. Typical contribution of the
extinction to the distance modulus uncertainty is also shown in 
Table \ref{tab:SEB}.

\subsubsection{Malmquist bias}
The Malmquist bias is a selection effect in flux limited samples. 
Due to the preferential detection of the most luminous objects
as a function of distance and limiting flux, at any distance 
there are always more 
faint objects being randomly scattered-in of the flux-limited
sample than bright
objects being randomly scattered-out of the sample. Therefore the
source mean
absolute magnitude at some large distance will be systematically
fainter than what expected due to the flux limit of the catalogue at
that distance.

The Malmquist bias for our flux limited low-$z$ calibrating sample was calculated
following the procedure given by \citet{Giraud1987}. In the first place, using the Luminosity Function for \hii\ galaxies  \citep{Chavez2014} we estimated the expected value of the
luminosity at any redshift as:
\begin{equation}
\langle L \rangle  = \frac{\int_{L_i}^{L_s}{L^{\alpha} L dL}}{\int_{L_i}^{L_s}{L^{\alpha}  dL}} ,
\end{equation}
where $L_i = 10^{39.7}$ is the lower limit of the luminosity function,
$L_s = 10^{42.5}$ is the upper limit and $\alpha = -1.5$ is the slope
\citep{Chavez2014}.

Subsequently, at each $z$ we calculate the luminosity  expected  when
we change the lower limit of the Luminosity Function to the  value
given by the flux limit at that redshift:
\begin{equation}
\langle L(z) \rangle  = \frac{\int_{L_l(z)}^{L_s}{L^{\alpha} L dL}}{\int_{L_l(z)}^{L_s}{L^{\alpha}  dL}} ,
\end{equation}
where the value of $L_l(z)$ can be calculated from:
\begin{equation}
 \log L_l(z) = \log f_l + 2\log(d_L[z, \mathbf{p}]) + 50.08 ,
\end{equation}
 where $\log f_l = -14.3$ is the flux limit of our low-$z$ sample and $d_L$ is the luminosity distance as function of $z$ and a set of 
 cosmological parameters $\mathbf{p}$.
 
Finally the bias is a function of the difference of the unbiased and biased  expected values of the luminosity and can be obtained as:
\begin{equation}
	b( \log L_{\mu}) = \frac{\sigma_0^2}{\sigma_{L_0}^2 + \sigma_0^2} (\log \langle L \rangle - \log \langle L(z) \rangle)
\end{equation}
where $\sigma_0$ is the 
dispersion of residuals of the \lsigb\ relation and $\sigma_{L_0}$ is
the  dispersion of the distribution of luminosities in the
sample. From the above equation the bias for a certain  distance modulus can be obtained as $b(\mu) = 2.5 b( \log L_{\mu})$.

The typical value of the Malmquist bias found for our low-$z$
calibrating sample is  $b(\mu) = 0.03$, extremely small  compared to
the other uncertainties.

\subsection{Gravitational Lensing Effects}
Details of the expected effect of
gravitational lensing on the distance modulus of high-$z$ standard
candles \citep[e.g.][and references therein]{Holz1998, Holz2005,
  Brouzakis2008} were given in  \citet{Plionis2011}. The basic
assumption used in developing a
correction procedure for this effect is that the magnification distribution resembles a
lognormal with zero mean (the mean flux of each source 
over all possible different paths is conserved, since lensing does not
affect photon numbers),
a mode shifted towards the de-magnified regime and a long tail towards
high magnification. This sort of distribution has been found in
analyses based on Monte-Carlo procedures and ray-tracing techniques \citep[cf.][]{Holz2005}.

Therefore most high-$z$ sources will be demagnified (will appear artificially fainter), inducing an apparently 
enhanced accelerated expansion, while a few will be highly magnified. The effect is obviously
stronger for higher redshift sources since the lower the redshift the smaller the optical 
depth of lensing. 

It is important to note that the effect of gravitational lensing is
not only to increase the distance modulus uncertainty, which is
proportional to the redshift, but 
also to induce a systematic shift of the mode of the
distance modulus distribution to de-magnified (fainter) values. 
These effects appear to be independent of the underlying cosmology and
the details of the density profile of cosmic structures \citep[eg.][]{Wang2002}.

A procedure, first suggested by
\citet{Holz2005}, to correct statistically for such an
effect was explained in detail  in \citet{Plionis2011}. The reader is referred to that work.
We apply this procedure to our analysis of the Hubble
expansion cosmological probe, using either \hii\ galaxies or SNIa, but
find minimal effects on the resulting cosmological parameter constraints.

\section{Cosmological Constraints}
A variety of observational probes have been developed through the years in order to
provide constraints on the cosmological parameters, which in turn
determine the specifics of the evolution of the Universe.
These probes may be divided in two general classes; {\em geometrical}
and {\em dynamical} and both use the redshift dependence of the comoving
distance to a source:
\begin{equation}\label{eq:Dc}
d_{C}(z)= \int_{0}^{z}{\frac{c dz^{'}}{H(z^{'})}}\;,
\end{equation}
where the Hubble function $H(z) [\equiv H_0 E(z)]$ is derived from the
first Friedman equation and $E(z)$ is given in the matter-dominated
era for a flat Universe with matter and DE, by:
\begin{equation}\label{eq:Ez}
E^2(z)= \left[ \Omega_{m,0}(1+z)^3 + \Omega_{w,0} (1+z)^{3y}
\exp\left( \frac{-3 w_a z}{z+1}\right) \right]
\end{equation}
with $y=(1 + w_0 + w_a)$.
The parameters $w_0$ and $w_a$ refer to the DE equation of
state, the general form of which is:
\begin{equation}
p_{w} =  w(z) \rho_{w} \;,
\end{equation}
with $p_{w}$ the pressure and $\rho_{w}$ the density of the
postulated DE fluid. 
Different DE models have been proposed and many are
parametrized using a Taylor expansion around the present epoch:
\begin{equation}
	w(a) = w_0 + w_a(1-a)\Longrightarrow w(z)=w_0+w_a \frac{z}{1 + z}\;,
\end{equation}
\citep[CPL model;][]{Chevallier2001, Linder2003, Peebles2003,
  Dicus2004, Wang2006}.
The cosmological constant is just a special case of DE, given
for $(w_0,w_a)=(-1,0)$, while the so called {\em quintessence} (QDE) models
are such that $w_a=0$ but $w_0$ can take values $\neq -1$.

Therefore, assuming a flat Universe ($\Omega_{m}+\Omega_{w}=1$), 
a negligible radiation density
parameter and the generic CPL DE EoS parametrisation, 
the most general set of cosmological parameters that is
necessary to be constrained in order to define the actual cosmological
model, is given by $\mathbf{p} = \{\Omega_{m,0}, w_0, w_a\}$. Note
that we do not include as a parameter the Hubble constant because, as
it will become clear further below, the dependence on $H_0$ is
factored out.
In what follows, we will consider two parametrisation of the DE EoS, assuming a flat Universe, i.e.,
\begin{enumerate}
\item QDE model with $\mathbf{p} = \{\Omega_{m,0}, w_0, 0\}$, and
\item CPL model with $\mathbf{p} = \{\Omega_{m,0}, w_0, w_a\}$
\end{enumerate}

The {\em geometrical} probes, which are independent of the underline
gravity theory, 
are used to probe the Hubble function
through the redshift dependence of the luminosity, $d_{L}(z)$, or the angular
diameter, $d_{A}(z)$, distance.

These methods utilize extragalactic sources for which their luminosity
is either known a priori (e.g. standard candles) or it can be estimated by
using a distance-independent observational parameter. Alternatively, they can  use cosmic
phenomena for which their metric size is known (e.g. standard rulers).
Then the cosmic
expansion history is traced via the luminosity distance $d_{L}(z)$, in
the first case, or the angular diameter distance $d_{A}(z)$, in the
second case. To date such observations probe
the integral of the Hubble expansion rate $H(z)$ either up to
redshifts of order $z\simeq 1.5$ (e.g., SNIa, BAO, clusters), or at
the redshift of recombination, $z_{rec}\sim 1100$ (CMB fluctuations).

{\em Dynamical} probes, on the other hand,  map the expansion history
 based on measures of the growth rate of cosmological perturbations
 and therefore depend on the theory of gravity \citep[cf.][and
   references therein]{Bertschinger2006, Nesseris2008,
   Basilakos2013}. Such methods are also confined to
relatively low redshifts, up to $z\simeq 1$. 

It is therefore clear
that the redshift range $1.5\lesssim z\lesssim 1000$ is not directly
probed to date by
any of the above cosmological tests, and as discussed in
\citet{Plionis2011} the redshift range $1.5\lesssim z\lesssim 3.5$ is
of crucial importance to constrain the DE EoS, 
since different DE models manifest their largest
deviations in this redshift range.
Therefore the fact that \hii\ galaxies can be observed relatively 
easily at such redshifts make them ideal and indispensable tools for
cosmological studies.  
Below we present the basics of the two {\em geometrical} probes that are
extensively used to constrain the DE EoS parameters.

\subsection{Standard Candle Probes}
As discussed previously, for standard candle probes we need to use the luminosity
distance of the sources tracing the Hubble expansion, given by
$d_{L}=(1+z) d_C$.  
For convenience, which will be understood below, we define a further
parameter, independent of the Hubble constant, by:
\begin{equation}
D_L(z,{\bf p})= (1+z) \int_{0}^{z}{\frac{dz^{'}}{E(z^{'},{\bf p})}}\;.
\end{equation}
i.e., $d_L=c D_L/H_0$.
Using the luminosity distance, as calculated 
from a set of cosmological parameters, $\mathbf{p}$, and 
the redshift, $z$, we can obtain the
`theoretical' distance modulus of a source as:
\begin{equation}
\mu_{th} = 5 \log d_L (\mathbf{p},  z) +25= 5 \log D_L (\mathbf{p},  z) + \mu_{0},
\label{eq:mu2}	
\end{equation}
where $\mu_{0}=42.384-5\log h$.
Therefore, to restrict a given set of cosmological parameters, we
define the usual $\chi^2$ minimisation function as:
\begin{equation}
\label{eq:mu22}	
\chi_{sc}^{2}({\bf p})=
\sum\limits_{i=1}^{N}\frac{[\mu_{\rm obs}(z_i)-\mu_{\rm th}(z_{i},{\bf p})]^2}
{\sigma_{\mu,i}^2}\;,
\end{equation}  
where $N$ is the total number of sources used, the suffix $sc$
indicates the standard candle probe and
$\mu_{\rm obs}(z_{i})$ and $\sigma^{2}_{\mu,i}$ are the distance 
moduli and the corresponding uncertainties at the observed redshift $z_{i}$.
Inserting the second equality of eq.(\ref{eq:mu2}) into eq.(\ref{eq:mu22}) 
we find after some simple algebra that 
\begin{equation}\label{eq:expand-xi2}
\chi_{sc}^2({\bf p})=A({\bf p})-2B({\bf p})\mu_0+C\mu_0^2\;,
\end{equation}
where
\begin{eqnarray}\label{eq:xi2-A-B-C}
 A({\bf p}) &=& \sum\limits_{i=1}^{N}\frac{[\mu_{\rm obs}(z_{i})-
5{\rm log}D_{L}(z_{i},{\bf p})]^2}{\sigma_{\mu,i}^2}\;,\nonumber \\
 B({\bf p}) &=&\sum\limits_{i=1}^{N}\frac{\mu_{\rm obs}(z_{i})-5{\rm log}D_{L}(z_{i},{\bf p})}
{\sigma_{\mu,i}^2}\;,\nonumber \\
 C &=& \sum\limits_{i=1}^{N} \frac{1}{\sigma_{\mu,i}^2}\;. \nonumber
\end{eqnarray}
Obviously for $\mu_{0}=B/C$ eq.(\ref{eq:expand-xi2}) 
has a minimum at 
\begin{equation}
{\tilde \chi}^{2}({\bf p})=A({\bf p})-\frac{B^2({\bf p})}{C}.
\end{equation}
Therefore, instead of using 
$\chi^{2}$ we now minimise ${\tilde \chi}^{2}$ 
which is independent of $\mu_{0}$ and thus of the value of the Hubble constant.
For more details concerning the above treatment 
the reader is referred to \citet{Nesseris2005}.
 	
\subsection{Standard Ruler Probes}
The first standard ruler probe is provided by 
the first peak of the CMB temperature perturbation spectrum, appearing
at $l_1^{TT}$, which refers to the angular scale of the sound horizon at the last
scattering surface, $\theta_{1}^{TT}\sim 1/l_1^{TT}$. Then by
calculating its comoving scale, $r_s(z_{rec})$, we can derive its angular
diameter distance by:
\begin{equation}
d_A(z_{rec}, {\bf p})=\frac{r_s(z_{rec}, {\bf p})}{\theta_1^{TT}}=\frac{d_C(z_{rec},
  {\bf p})}{1+z_{rec}}\;.
\end{equation}
Since the above equation is model dependent, through the CMB physics
determination of $r_s$, a model independent parameter has been
defined, the so-called {\em shift parameter} \citep{Bond1997,
  Nesseris2007}, which is the ratio of the position of the first peak
to that of a reference model, and for spatially flat models it is
given by:
\begin{equation}\label{shiftparameter}
R({\bf p})=\sqrt{\Omega_{m,0}}\int_{0}^{z_{rec}} \frac{dz}{E(z, {\bf p})}\,.
\end{equation}
The observationally measured shift parameter, according to the recent
{\em Planck} data \citep{Shafer2014} is $R=1.7499\pm 0.0088$ at the 
redshift of decoupling (viz. at the last scattering surface, $z_{rec}=1091.41$). 
At this point we would like to remind the reader that 
when dealing with the CMB
shift parameter we need to include also the radiation density
term in the $H(z)$ function since at recombination it amounts to $\sim
23\%$ of the matter density ($\Omega_{r,rec}\simeq 0.23 \Omega_{m,rec}$)
and therefore cannot be ignored. The final minimisation function is:
\begin{equation}
\label{eq:cmb}	
\chi^{2}_{\rm CMB}({\bf p})= \frac{[R({\bf p})-1.7499]^2}{0.0088^2}\;.
\end{equation}  

The second standard ruler probe that we use
is the Baryonic Acoustic Oscillation (BAO) scale, a feature produced in the last
scattering surface by the competition between the pressure of the
coupled baryon-photon fluid and gravity. The resulting sound
waves leave an overdensity
signature at a certain length scale of the matter distribution. This
length scale is related to the comoving distance that a sound
wave can travel until recombination and in practice it manifests
itself as a feature in the correlation function of galaxies on large
scales ($\sim 100 \;h^{-1}$ Mpc).
In recent years, measurements of the BAO have proven extremely useful as 
a ``standard ruler''.
The BAOs were clearly identified, for the first time,
as an excess in the clustering pattern of the SDSS luminous red galaxies
\citep{Eisenstein2005}, and of the 2dFGRS galaxies \citep{Cole2005}.
Since then a large number of dedicated surveys have been used to
measure BAOs, among which the  WiggleZ Dark Energy Survey
\citep{Blake2011}, the 6dFGS \citep{Beutler2011} and
the SDSS Baryon Oscillation Spectroscopic Survey (BOSS) of SDSS-III
\citep{Eisenstein2011,Anderson2014,Aubourg2015}.

In the current paper we utilise the results 
of \citet[][see their Table 3]{Blake2011}
which are given in terms of the acoustic parameter $A(z)$,
first introduced by \citet{Eisenstein2005}:

\begin{equation}\label{defBAOA}
A({z_i,\bf p})=\frac{\sqrt{\Omega_{m,0}}}{[z_i^{2} E(z_{i}, {\bf p})]^{1/3}}
\left[\int_{0}^{z_i} \frac{dz^{'}}{E(z^{'}, {\bf p})} \right]^{2/3}
\end{equation}
with $z_i$ the redshift at which the signature of the acoustic oscillations has been
measured. The corresponding minimisation function is given by 
\begin{equation}
\label{eq:BAO}	
\chi^{2}_{\rm BAO}({\bf p})= \sum_{i=1}^{6}\frac{[A(z_i,{\bf p})-A_{obs,i}]^2}{\sigma_i^2}\;.
\end{equation}  
where $A_{obs,i}$ are the observed $A_i$ values at six different redshifts, $z_i$,
provided in Blake et al. (2011).

\subsection{Joint Analysis of Different Probes}
In order to place tight constraints on the corresponding
parameter space of the DE EoS, the cosmological 
probes described previously must be
combined through a joint likelihood analysis, given by the product
of the individual likelihoods according to:
\begin{equation}
{\cal L}_{\rm tot}({\bf p})= \prod_{i=1}^{n} {\cal L}_i({\bf p})
\label{eq:overalllikelihood}
\end{equation}
where $n$ is the total number of cosmological probes
used\footnote{Likelihoods are normalised to their maximum
values.}. This
translates to an addition for the corresponding joint total $\chi^2_{\rm tot}$
function:
\begin{equation}
 \chi^{2}_{\rm tot}({\bf p})=\sum_{i=1}^{n}\chi^{2}_i({\bf p})
\;.
\label{eq:overalllikelihoo}
\end{equation}
In our current analysis we sample the cosmological
parameter space with the following resolution: 
$\delta\Omega_{m,0}=0.001$, $\delta w_0=0.003$ and
$\delta w_a=0.016$. Also, the reported uncertainties for each
unknown parameter of the vector ${\bf p}$ are estimated after marginalising one
parameter over the other, such that $\Delta \chi^{2}(\le 2\sigma)$.
Note however that in order to appreciate the possible degeneracy among
the different fitted parameters one must inspect the two-dimensional
likelihood contours.
\begin{figure}
  \mbox{\epsfxsize=8.2cm 
\epsffile{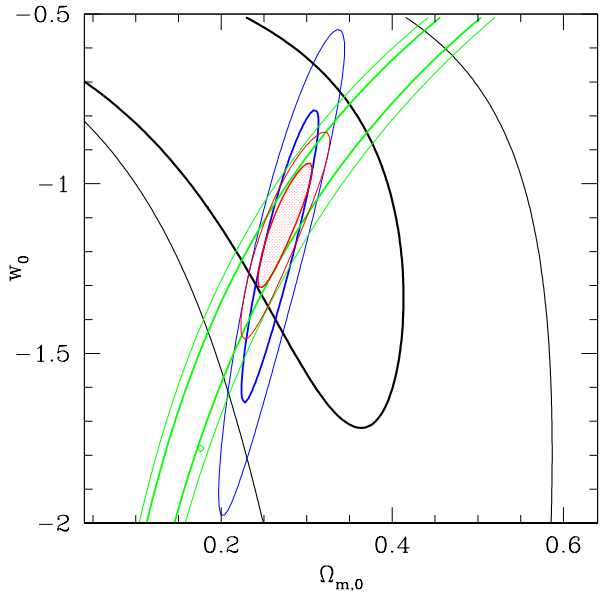}} 
\caption{Likelihood contours for $\Delta
\chi^2=\chi^{2}_{\rm tot}-\chi^{2}_{\rm tot,min}$ equal to 2.32 and 6.18
corresponding to the 1$\sigma$ and 2$\sigma$ 
confidence levels in the $(\Omega_{m,0},w)$ plane. 
Results based on the \hii\ galaxies are shown in black, on the 
CMB shift parameter (green) and on BAO (blue) while the joint contours
are shown in red.}
\label{fig:01}
\end{figure}

\begin{table*}
\caption{Cosmological parameters from the joint analysis of different
  combinations of probes and for both parameterisations of the DE EoS.}
\label{tab:02}
\tabcolsep 12pt
\begin{tabular}{cccccc} \hline
Probes        & $\Omega_{m,0}$ & $w_0$ & $w_a$ & $\chi^2_{min}$ & df
\\ \hline
& \multicolumn{5}{c}{QDE parameterisation} \\
BAO/CMB       & 0.274$\pm0.0145$&$-1.109\pm0.082$& 0 & 1.036 & 6 \\
HII/BAO/CMB   & 0.278$\pm0.0143$&$-1.088\pm0.080$& 0 & 213.85 & 162 \\
SNIa/BAO/CMB  & 0.287$\pm0.0130$&$-1.034\pm0.056$& 0 & 563.68 & 586 \\
& \multicolumn{5}{c}{CPL parameterisation} \\
BAO/CMB       & 0.278& $-1.052\pm0.083$&$-0.112\pm0.35$ & 1.087 & 6 \\
HII/BAO/CMB   & 0.278& $-0.992\pm0.084$&$-0.368\pm0.38$ & 213.72 & 162 \\
SNIa/BAO/CMB  & 0.278& $-0.983\pm0.057$&$-0.304\pm0.28$& 563.90 &
586 \\ \hline
\end{tabular}
\end{table*}

\begin{figure*}
\mbox{\epsfxsize=17.2cm 
\epsffile{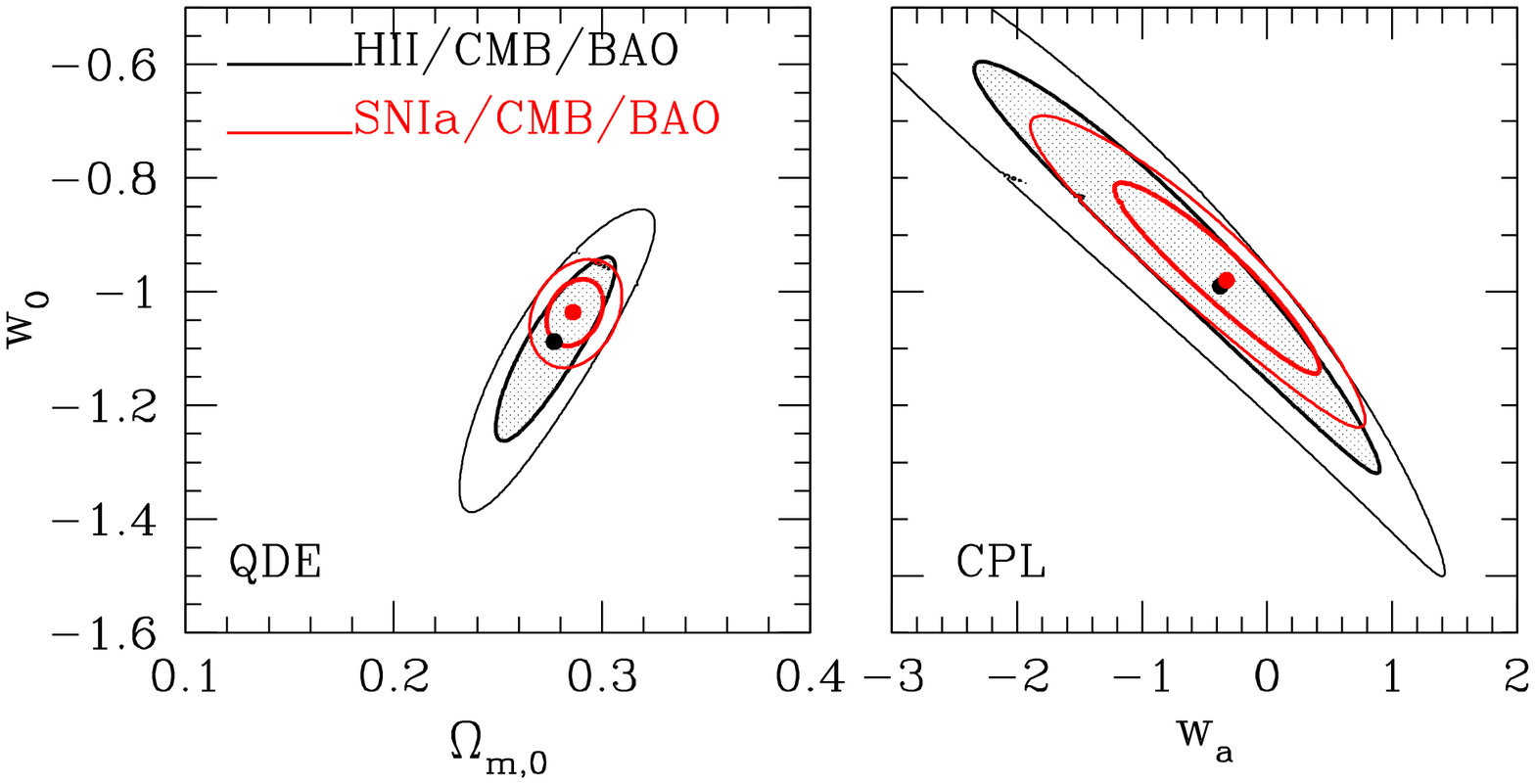}} 
\caption{Comparison of the joint likelihood contours of the
  HII/CMB/BAO (black contours) and of the SNIa/CMB/BAO 
  (red contours) probes.
{\it Left Panel:} QDE dark energy equation of state parametrisation.
{\it Right Panel:} CPL dark energy equation of state parametrisation
using $\Omega_{m,0}=0.278$ as a prior.}
\label{fig:02}
\end{figure*}

\subsection{Results of the Joint Analysis}
As discussed earlier our present sample of \hii\ galaxies is dominated by the very-low
redshift regime ($z<0.15$) as it contains only a small number of high-$z$
sources; therefore the cosmological constraints that can be imposed 
are very weak \citep[see][]{Terlevich2015}. Nevertheless by joining the
\hii\ galaxy analysis with other cosmological probes we can further test the
effectiveness of using \hii\ galaxies as alternative tracers of the
Hubble expansion.
To this end we will present and compare our results of the joint analysis 
but using as standard candles separately our \hii\
galaxies and  the SNIa. 

In Figure \ref{fig:01} we present the 1$\sigma$ and 2$\sigma$ likelihood 
contours in the $(\Omega_{m,0}, w)$ plane for the following
probes: \hii\ galaxy Hubble relation (black contours), 
CMB shift parameter (green) and BAO (blue), whereas with red we
present the result of the joint analysis. The solution provided by the
\hii\ galaxy Hubble relation probe has been shown to be consistent with that
of the SNIa, albeit leaving mostly unconstrained the QDE free
parameters \citep{Terlevich2015}. However, 
the joint analysis reduces dramatically the solution space, providing quite
stringent constraints on the two QDE parameters. Even with the current
very broad \hii\-galaxy likelihood contours, the joint  HII/BAO/CMB
analysis increases the Figure of Merit (FoM) by 13\% with respect to that of the BAO/CMB
joint analysis alone.
 
In order to compare the performance of the \hii\ galaxies (as they stand
today in our sample of only 25 high-$z$ sources) with that of the {\em Union2.1}
SNIa, we display in Figure \ref{fig:02} the joint likelihood contours for 
HII/BAO/CMB$_{\rm shift}$ (black contours),  
and SNIa/BAO/CMB$_{\rm shift}$ (red contours) probes for both DE EoS parameterisations.
Note that in the case of the CPL analysis we impose an {\em a priori}
value for the cosmological matter density parameter, 
$\Omega_{m,0}=0.278$, and allow the two DE EoS parameters,
$w_0$ and $w_a$ to vary.

A first observation is that both joint analyses, based either on \hii\
galaxies or SNIa, provide consistent results for both DE equation of
state parameterisations, 
although (as expected) the SNIa rate better since
the SNIa sample is much larger and their median redshift is
significantly higher than that of our preliminary \hii\ galaxy sample.
For the QDE case, the broad
\hii\ galaxy likelihood contours and the corresponding extensive
parameter degeneracy is reduced significantly with the joint 
HII/BAO/CMB analysis, while
the degeneracy appears to disappear with the SNIa/BAO/CMB analysis. 
As expected for the more
demanding CPL parametrisation the degeneracy between $w_0$ and $w_a$
is present in both sets of joint analyses.
However, what is particularly interesting is that
for the CPL model the two joint analyses
provide the same minimum, as can be seen also in Table \ref{tab:02},
where we list the resulting cosmological parameters and their
uncertainties for the different combinations of cosmological
probes.

It is very encouraging that even with the current \hii\ galaxy pilot sample, the
combined analysis of the \hii\ data with BAOs
and the CMB shift parameter provides constraints on the cosmological
parameters which are in agreement with those of the joint
SNIa/BAOs/${\rm CMB}_{\rm shift}$.

We plan to considerably increase the current sample of high-$z$ \hii\
galaxies (see next section) which together with other future cosmological data, based
for example on {\em Euclid}, will improve significantly the relevant
constraints (especially on $w_{a}$) and thus the validity of
a running EoS parameter, namely $w(z)$,
will be effectively tested.

\section{Monte-Carlo Simulations}
In order to predict the effectiveness of using high-$z$ \hii\ galaxies to
constrain the DE EoS, we have performed an
extensive series of Monte-Carlo simulations with which we assess our
ability to recover the input parameters of an {\em a priori} selected cosmological
model, in our case that of the concordance cosmology ($\Omega_{m, 0},
w_0, w_a)=(0.28, -1, 0)$.
We distribute different numbers of mock high-$z$ \hii\ galaxies
in redshift according to the observational constraints of
the adequate, for our purpose, instruments and telescopes (in this case
the VLT-KMOS spectrograph at ESO\footnote{As we prepared this work, we
  have also procured some 25 high-z \hii\ galaxies data with MOSFIRE
 at Keck. A paper is in preparation.}). The range of the available near IR bands
for this instrument are shown in Table \ref{tab:03}, as well as the corresponding
redshift ranges within which either the H$\alpha$  or [OIII] emission lines can
be observed. There are practically 4 independent redshift ranges that
can be sampled 
centered at $\langle z \rangle \simeq 0.8, 1.4, 2.3$ and 3.3, and 
these are the redshift ranges where we will distribute
our mock high-$z$ \hii\ galaxies\footnote{Note that other studies that
  present simulations of the constraints provided by future
  high-$z$ tracers of the Hubble expansion do not always take into
  account the limited redshift intervals that can be observationally probed
  \citep[cf.][]{Scovacricchi2016}.}. Since the IR bands window function are
clearly not top-hat, we model the distribution of redshifts, within
each $z$-window, by a Gaussian with mean and standard deviation given in
Table \ref{tab:03}.

\begin{table}
\caption{\bf The KMOS FWHM sensitivities and redshift windows}
\label{tab:03}
\tabcolsep 2pt

\begin{tabular}{ccccc} \hline
Band & $\lambda$/nm & H$\alpha$ $z$-window & [OIII] $z$-window & Exp.time (sec) \\
& & & & $S/N\simeq 25$ \\
J & 1175$\pm 40$ & 0.79$\pm 0.026$ & 1.35$\pm 0.046$ & 1800 \\
H & 1635$\pm 65$ & 1.49$\pm 0.060$ & 2.26$\pm 0.090$ & 1500 \\
K & 2145$\pm 65$ & 2.26$\pm 0.070$ & 3.28$\pm 0.100$ & 2100 \\ \hline
\end{tabular}
The width of the wavelength coverage includes only the region
  with sensitivity higher than 50\% of the band peak sensitivity.
\end{table}

The Monte-Carlo simulation procedure that we follow entails assigning
to each mock \hii\ galaxy the ideal distance modulus for the
selected cosmology and an uncertainty which is determined by the
expected distribution of luminosity and flux errors that enter in  
the relation (\ref{eq:mu1}). We then transform these errors in a
distance modulus error distribution 
and use this distribution to assign randomly errors to each high-$z$ mock \hii\
galaxy. The mean distance modulus uncertainty is thus derived from
propagating the mean velocity dispersion and flux errors via
eq.(\ref{eq:mu1}), i.e.:
\begin{equation}
\sigma_\mu= 2.5 \left( \log\sigma^2 \sigma_a^2 +
a^2\sigma^2_{\log\sigma}+\sigma_b^2 +\frac{\sigma_f^2}{\ln(10)^2
  f^2}\right)^{1/2}
\end{equation}
where $a$ and $b$ are the slope and intercept of the
$L_{H\beta}-\sigma$ relation, $\sigma_a$ and $\sigma_b$ are the
corresponding uncertainties of the fit, while $f$ and $\sigma_f$ are
the $H\beta$ line flux and its uncertainty.
Assuming a flux uncertainty of $\lesssim 10\%$ \citep[as indeed we
find for the three $z\gtrsim 1.5$ \hii\ galaxies we observed with X-SHOOTER; see][]{Terlevich2015} and the uncertainties of our
$L_{H\beta}-\sigma$ relation, we obtain a mean 
$\langle \sigma_\mu\rangle \simeq 0.6$ mag, slightly lower than the
measured values of
our low-$z$ sample ($\langle \sigma_\mu\rangle \simeq 0.7$ mag). 

The available high-$z$ \hii\ galaxy data from the literature (as well as our own data) indicate a large
dispersion of the distance modulus uncertainty and therefore, for the
purpose of our simulations, we will assume a Gaussian uncertainty
distribution with mean $\langle \sigma_\mu\rangle \simeq 0.6$ mag
and a standard deviation of $\sigma_{\sigma}\simeq 0.24$. Obviously,
the outcome of the simulations are sensitive to the error distribution
and the results presented here are intended as indicative of the potential
of our approach.

\subsection{Results of simulations}
\begin{figure}
\mbox{\epsfxsize=8.9cm 
\epsffile{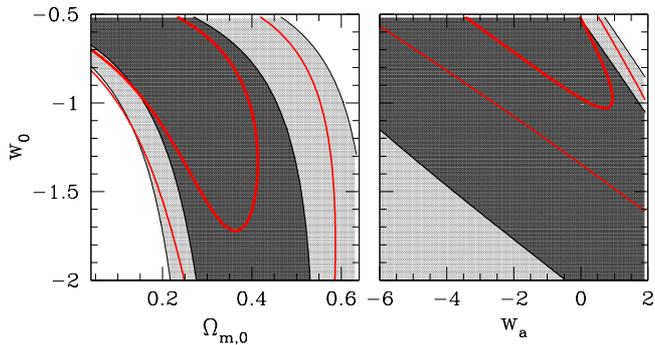}}
\caption{Likelihood contours 
corresponding to the 1$\sigma$ and 2$\sigma$ 
confidence levels for our \hii\
galaxy sample but using the ideal concordance cosmology distance
moduli (grey-scale contours). In red we show the corresponding true
constraints of our current sample.  {\em Left Panel:} QDE parametrisation. {\em
  Right Panel:} CPL parametrisation with $\Omega_{m,0}=0.278$.}
\label{fig:03}
\end{figure}

In order to test the effectiveness of our procedure, as a starting
point, we assign to each of the 156 \hii\ galaxies and GEHR of our high-quality 
velocity dispersions observational sample \citep{Chavez2014, Terlevich2015} 
the ideal distance modulus and the actual observed
uncertainty. We then perform our usual
$\chi^{2}$ minimisation procedure and derive the cosmological
constraints, shown in Figure \ref{fig:03} as greyscale contours.
We also overplot the corresponding true
observational constraints of the same \hii\ galaxy sample, which
are statistically consistent with the {\em ideal} case (more so for
the QDE parametrization).
If for the {\em ideal} distance modulus case we assign to each source
the model observational uncertainties, discussed previously, we obtain
similar constraints as in the true uncertainties case but with
slightly higher FoM, by a factor of $\lesssim 2$.

For our tests we will consistently estimate the increase of the 
current FoM, based on the 156 \hii\ galaxies and GEHR of our sample using the
{\em ideal} distance moduli 
with that provided when we add 
different numbers of high-$z$ \hii\ galaxies, distributed in the
redshift ranges shown in Table \ref{tab:03}. This exercise will be presented
for both the QDE and CPL parameterisations of the DE EoS.
Note that the distribution of numbers of the mock \hii\ galaxies at the different
redshift ranges could also affect the results in the sense that
different cosmological models show the largest deviations from the
concordance model at different redshifts \citep[eg., Fig.1 of][]{Plionis2011}. 
After a trial and error procedure we found that an optimal
distribution of the fractions of the total number of high-$z$ \hii\
galaxies in the 4 available redshift ranges, shown in Table 3, is 0.2,
0.2, 0.3 and 0.3 (from the lowest to the highest redshift
range). However, the case of equal fraction among the different redshifts 
 provide  similar results.

We performed 100 Monte-Carlo realisations for each selected number of
mock high-$z$ \hii\ galaxies, and the aggregate results are presented in
Figure 4, in the form of the ratio between the simulation FoM and that
of our current sample of \hii\ galaxies as a function of the number of
mock high-$z$ \hii\ galaxies. Thus what is shown is the factor by which
the FoM increases with respect to its current value. This factor
increases linearly with $\rm N_{\rm HII}$ providing the following
rough analytic expressions:
$$F_{\rm QDE}\simeq 0.015 N_{\rm HII}+ 1.72 \;\;\;\;{\rm and} \;\;\;\; F_{\rm CPL}\simeq
0.004 N_{\rm HII}+ 1.51$$
which means that for the very realistic near future expectations of observations of $\sim 500$
high-$z$ \hii\ galaxies, we predict a $\sim$ ten-fold increase of the current
FoM for the QDE parametrisation and $\sim$ four-fold increase of the
corresponding FoM for the CPL parametrisation, within the limits of
the parameters shown in Figure \ref{fig:04}.

As an example, we present in Figure \ref{fig:05} the results of one simulation of 500 high-$z$ mock
\hii\ galaxies both for the QDE and CPL parameterisations of the DE
EoS (grey-scale contours), which can be compared with
the constraints of our current sample (but using for consistency the
ideal distance moduli).

\section{Feasibility of the project and Future Work}
The realisation of this project relies on two main prerequisites; finding an
adequate number of high-$z$ \hii\ galaxy targets and being able to 
observe them using a reasonable amount of observing time.

\begin{figure}
\mbox{\epsfxsize=8.2cm 
\epsffile{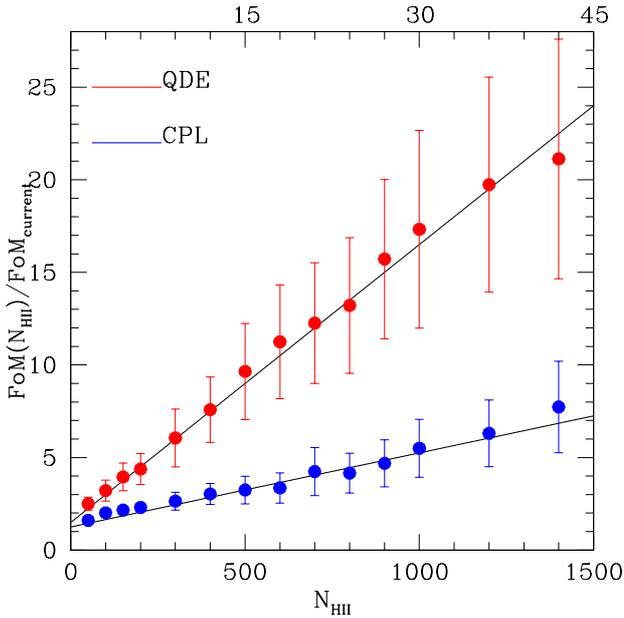}}
\caption{The factor by which the FoM of the QDE and CPL EoS constraints 
increases with respect to its current value (based on the observed 25
high-$z$ HII galaxies) as a function of the number of
mock high-$z$ \hii\ galaxies. The FoM has been estimated within the limits of the parameters
shown in Figure \ref{fig:03}. The red and blue points correspond to the
QDE and CPL parameterisations of the DE EoS, 
respectively. The solid black lines are
the linear fits to the corresponding coloured curves. The scale at the
top gives the number of 10m class telescope nights needed in order 
to observe 500, 1000 and 1500 objects as 15, 30 and 45 nights respectively.}
\label{fig:04}
\end{figure}

\begin{figure}
\mbox{\epsfxsize=8.9cm 
\epsffile{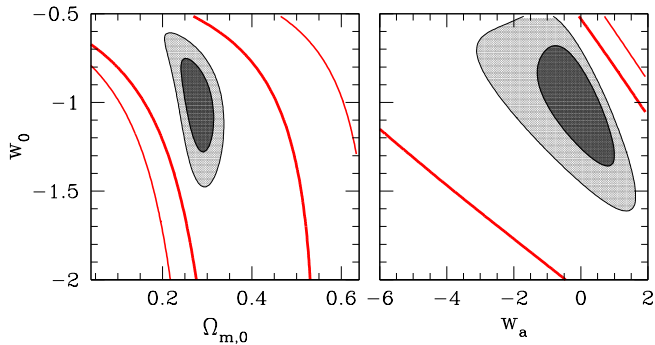}}
\caption{Likelihood contours 
corresponding to the 1$\sigma$ and 2$\sigma$ 
confidence levels for our \hii\
galaxy sample but adding 500 high-$z$ mock \hii\ galaxies (grey-scale
contours). In red we show the corresponding current
constraints (ie., without the high-$z$ mock \hii\ galaxies). We consistently
use the {\em ideal} distance moduli of the concordance cosmology. {\em Left Panel:} QDE parametrisation. {\em
  Right Panel:} CPL parametrisation using $\Omega_{m,0}=0.278$.}
\label{fig:05}
\end{figure}

To this end, we compiled a sample of objects searching the literature for high-$z$
\hii\ galaxy candidates that we define as compact emission line  systems
with either $W$(H$\alpha)>200$ \AA\ and $W$[OIII]$\lambda 5007 > 200$ \AA\  or
with $W$(H$\beta)>50$ \AA\ and FWHM$<150$ \AA\  and with $z>1.2$. We have found
up to now more than 500 candidates in about 20 high galactic latitude fields
 (Gonz\'alez-Mor\'an et al.~in preparation).
To estimate the feasibility of our project we calculated the time it
could take to observe the whole sample. For this estimate we have
assumed the use of IR spectroscopic facilities with resolution R
larger than 4000 in 10m class telescopes and with multiplexing
capability. These facilities are at present only two, KMOS at the VLT
and MOSFIRE at Keck. We have used the KMOS Exposure Time Calculator to
estimate the  time needed to obtain a S/N 25 or larger in either
H$\alpha$ or [OIII]$\lambda$5007 for the faintest objects in our list
and combine this estimate with their surface density at $z\sim 2.3$.
The  typical exposure times are about 3 hours per field. Each search
field is typically populated by 25 objects with about 8 to 15
simultaneously inside either the KMOS or MOSFIRE field of view. Thus
the number of objects that can be observed in a 10 hours night ranges
from 24 to 45, 
therefore about 15 observing nights would be needed to observe 500 \hii\ galaxies.
This estimate is shown in the upper scale of Figure \ref{fig:04}.

\section{Conclusions}
We have used the Hubble relation of  \hii\ galaxies 
in a joint likelihood
analysis with the BAO and CMB cosmological probes 
with the aim of testing the consistency of the derived
cosmological constraints with those of the joint
SNIa/BAO/CMB analysis. This results in two important conclusions:
\begin{itemize}
\item The FoM of the QDE EoS constraints, provided by the joint
  HII/BAO/CMB analysis, was found to be larger by 13 percent than
  those provided by the BAO/CMB joint analysis, even with the very small
  sample of only 25 high-z HII galaxies.
\item Both the QDE and CPL EoS constraints of the HII/BAO/CMB and of
  the SNIa/BAO/CMB joint analyses are in excellent consistency with
  each other, although (as expected) the SNIa probe still provides a
  significantly larger FoM.
\end{itemize}

We have also performed Monte-Carlo simulations tailored to the specific
uncertainties of the \lsigb\ relation and to the  technical
instrumental requirements of KMOS/VLT (and instruments like it). They
address the important question of what is the expected increase of the
FoM as a function of the number of high-z \hii\ galaxies in the
redshift windows accessible. Our previous
simulations \citep[cf.][]{Plionis2011} did not take into account
 the specific error budget of our \lsigb\ relation, or the
characteristics of the instruments available and of the accessible
redshifts.  
We would like to add that cosmological analyses, like the one
presented in this work, demands a
thorough understanding of the interplay between observational random
and systematic errors and biases, for which mock catalogues are an
essential tool.

\section*{Acknowledgements}
We are thankful to an anonymous referee for careful and constructive comments on the manuscript.
RC, RT, ET  and MP are grateful to
the Mexican research council (CONACYT) for supporting this research
under studentship 224117 and grants 263561, CB-2005-01-49847, CB-2007-01-84746 and
CB-2008-103365-F.  
SB acknowledges support by the Research
Center for Astronomy of the Academy of Athens in the context of the
program {\it ``Tracing the Cosmic Acceleration''}. MP acknowledges the
hospitality of the KAVLI Institute for Cosmology in Cambridge, where
this work was completed. 

\bibliography{bib/bibpaper2016}
\label{lastpage}

\end{document}